\documentclass[
aps,
prd,
reprint,
superscriptaddress,
nofootinbib,
]{revtex4-2}

\usepackage{graphicx}
\usepackage[usenames,dvipsnames,table,xcdraw]{xcolor}
\usepackage{xspace}
\usepackage{amsmath,amsfonts,amssymb,amsthm,bm}
\usepackage[varg]{txfonts} 
\usepackage{mathtools}
\usepackage[utf8]{inputenc}
\usepackage[normalem]{ulem}
\usepackage{makecell}
\usepackage{multirow}
\usepackage{float}
\usepackage{adjustbox}
\usepackage[caption=false]{subfig}
\usepackage{rotating}
\usepackage{dcolumn}
\usepackage{enumitem}
\usepackage{natbib}
\usepackage[htt]{hyphenat}
\xdefinecolor{mylinkcolor}{rgb}{0,0,0.7}
\usepackage[
	bookmarksnumbered, bookmarksopen, bookmarksopenlevel=2,
	breaklinks=true,
	colorlinks=true, filecolor=mylinkcolor, citecolor=mylinkcolor,
	linkcolor=mylinkcolor, urlcolor=mylinkcolor, menucolor=mylinkcolor,
]{hyperref}

\usepackage{setspace}
\usepackage[hang]{footmisc} 
\allowdisplaybreaks

\graphicspath{ {fig/} }

\newcommand{\AEI}{\affiliation{Max Planck Institute for Gravitational Physics (Albert Einstein Institute), Am M\"uhlenberg 1, Potsdam 14476, Germany}}
\newcommand{\Maryland}{\affiliation{Department of Physics, University of Maryland, College Park, MD 20742, USA}}
\newcommand{\UoN}{\affiliation{Nottingham Centre of Gravity \& School of Mathematical Sciences, University of Nottingham, University Park, Nottingham NG7 2RD, United Kingdom}}

\begin{document}

\title{
Joint inference of line-of-sight acceleration and orbital eccentricity\\in neutron star--black hole binaries
}

\author{Lorenzo Pompili}
\email{lorenzo.pompili@nottingham.ac.uk}
\UoN

\author{Aldo Gamboa}
\AEI

\author{Alessandra Buonanno}
\AEI
\Maryland


\begin{abstract}
A line-of-sight acceleration (LOSA) of a compact-binary center of mass, imparted for example by a nearby tertiary perturber, imprints a Doppler modulation on the gravitational-wave signal and provides a single-event diagnostic of dynamical formation environments. Waveform-modeling systematics---missing higher-order modes, spin precession, or orbital eccentricity---can mimic or mask a non-zero LOSA, making waveform accuracy a leading concern for LOSA inference. We implement LOSA corrections directly in the time domain as a remap of the strain: the treatment applies uniformly to all mode content, spin precession, and orbital eccentricity, and integrates naturally with time-domain inspiral--merger--ringdown models that best capture these effects. We deploy it in the \texttt{SEOBNRv6EHM} (aligned-spin eccentric) and \texttt{SEOBNRv5PHM} (precessing quasi-circular) models, and validate the implementation on simulated signals; we find that \texttt{SEOBNRv6EHM} recovers LOSA correctly on both eccentric and spin-precessing injections, while \texttt{SEOBNRv5PHM} yields a spurious LOSA measurement on eccentric signals. Motivated by the eccentricity hints in the neutron-star--black-hole (NSBH) event GW200105\_162426 and the favorable low-mass regime of these sources, we jointly infer LOSA and orbital eccentricity for the five NSBH events in the LIGO--Virgo--KAGRA catalog, reporting the first LOSA constraints on three of them. All five events are consistent with a vanishing LOSA, $\Gamma \equiv a_\parallel/c = 0$; for GW200105\_162426, the joint $(\Gamma, e)$ posterior nonetheless disfavors both $\Gamma$ and $e$ being zero simultaneously at 90\% credibility, supporting the eccentricity hints reported in previous analyses.
\end{abstract}

\date{\today}

\maketitle


\section{Introduction}
\label{sec:intro}

The detection of gravitational waves (GWs) from compact-binary coalescences by the LIGO--Virgo--KAGRA (LVK) interferometers~\cite{TheLIGOScientific:2014jea, VIRGO:2014yos, KAGRA:2020tym} has opened a new observational window on the astrophysics of black holes (BHs) and neutron stars (NSs)~\cite{LIGOScientific:2016aoc, LIGOScientific:2018mvr, LIGOScientific:2020ibl, LIGOScientific:2021usb, Kagra:2021vkt, LIGOScientific:2025slb, LIGOScientific:2026wfs}.
With nearly $300$ events reported through the second part of the fourth observing run, the focus of GW astronomy is broadening from individual detections to the astrophysical origin of compact-binary mergers~\cite{LIGOScientific:2026ctl}. Among the source properties carried by the GW signal, orbital eccentricity is a particularly clean discriminator of formation channels. Binaries formed through isolated stellar evolution circularize efficiently via GW emission long before entering the sensitive band of ground-based detectors~\cite{Peters:1963ux,Peters:1964zz}; by contrast, binaries assembled dynamically can retain measurable eccentricity in the frequency range to which detectors are sensitive~\cite{Wen:2002km, OLeary:2008myb, Samsing:2013kua, Samsing:2017xmd, Rodriguez:2018pss, Zevin:2018kzq, Samsing:2020tda, Tagawa:2020jnc, Zevin:2021rtf, Chattopadhyay:2023pil, Stegmann:2025clo, Rozner:2026jtj}.

Recent years have seen significant progress in modeling eccentric compact-binary signals across the full inspiral--merger--ringdown (IMR) regime. This includes the development of effective-one-body (EOB) models within the \texttt{SEOBNR}~\cite{Ramos-Buades:2021adz, Khalil:2021txt, Gamboa:2024hli, Gamboa:2024imd, Gamboa:2026jht} and \texttt{TEOBResumS}~\cite{Chiaramello:2020ehz, Nagar:2021gss, Nagar:2024oyk, Albanesi:2025txj} families, time- and frequency-domain phenomenological models~\cite{Planas:2025feq, Ramos-Buades:2026kbq}, eccentric spin-precessing post-Newtonian (PN) inspiral-only models~\cite{Klein:2018ybm, Morras:2025nlp, Morras:2026fho}, and non-spinning numerical-relativity (NR) surrogates~\cite{Nee:2025nmh, Islam:2026blk}.
In this work we focus on \texttt{SEOBNRv6EHM}~\cite{Gamboa:2026jht}, an IMR EOB model for aligned-spin binary black holes (BBHs) in generic planar orbits, and the most NR-faithful eccentric aligned-spin model currently available~\cite{Gamboa:2026jht}. The model is also computationally efficient, enabling analyses of long-duration signals~\cite{Pompili:2026yxq}.

Parameter-estimation analyses with these models~\cite{Romero-Shaw:2022xko, Gupte:2024jfe, Morras:2025xfu, Planas:2025plq, Planas:2025jny, Jan:2025fps, Kacanja:2025kpr, Gupte:2026whi, Malagon:2026uev, Pompili:2026yxq} have identified a small set of potentially eccentric BBH candidates among the LVK detections through the first part of the fourth observing run (O4a), together with the neutron-star--black-hole (NSBH) event GW200105\_162426, with Bayes factors favoring eccentricity over a quasi-circular hypothesis. Their interpretation is, however, complicated by the eccentricity--spin-precession degeneracy~\cite{Romero-shaw:2022fbf, CalderonBustillo:2020xms}, by non-Gaussian noise artifacts~\cite{Gupte:2024jfe, Gupte:2026whi}, by sensitivity to the choice of eccentricity prior~\cite{Morras:2025xfu, Planas:2025jny, Clarke:2026cuw}, and by waveform systematics~\cite{Jan:2025zcm, Malagon:2026uev, Pompili:2026yxq}.

The same dynamical environments that can produce non-zero orbital eccentricity can also impart a center-of-mass acceleration on the binary. This may be sourced, for example, by the tertiary in a comparable-mass hierarchical triple, or by the orbital motion of a stellar-mass binary within the deep potential of a supermassive black hole (SMBH), as in a nuclear star cluster or an active galactic nucleus (AGN) disk~\cite{Yunes:2010sm, Meiron:2016ipr, Inayoshi:2017hgw, Robson:2018svj, Toubiana:2020drf, Samsing:2024syt, Hendriks:2024zbu}. A non-zero line-of-sight acceleration (LOSA), $a_\parallel$, imprints a time-dependent Doppler modulation on the GW signal~\cite{Bonvin:2016qxr, Vijaykumar:2023tjg}, which provides an event-by-event probe of the perturber's mass and separation. LOSA is one of several environmental effects that can leave imprints on the GW signal~\cite{Barausse:2014tra}, and stands out as one of the few potentially accessible at the single-event level by current ground-based detectors. Eccentricity and LOSA therefore constitute a complementary pair of formation-channel diagnostics, both arising from interactions between the binary and its environment, and most informative when measured jointly~\cite{Zwick:2025qzv, Tagawa:2025tfd, Takatsy:2025bfk}. Their leading-order phase corrections, however, share a similar frequency scaling ($f^{-34/9}$~\cite{Krolak:1995md} versus $f^{-13/3}$~\cite{Bonvin:2016qxr}) and exhibit a partial degeneracy in the inspiral~\cite{Bhat:2025lri, Pathak:2026cik, Roy:2026mco}.

The theoretical treatment of LOSA in compact-binary waveforms is well developed. Ref.~\cite{Yunes:2010sm} first studied the effect of a center-of-mass acceleration on GW waveforms, in the context of LISA observations of extreme mass-ratio inspirals (EMRIs) perturbed by a secondary SMBH, showing that, for sufficiently close perturbers, higher time-derivatives of the line-of-sight velocity also become measurable, allowing the perturber's mass and distance to be inferred independently. Ref.~\cite{Bonvin:2016qxr} derived the analogous leading-order frequency-domain amplitude and phase corrections for quasi-circular, comparable-mass binaries, showing that the effect enters at $-4$PN relative to the leading quadrupole order, and is therefore enhanced at low frequencies and for low-mass binaries whose signals spend long in band. Ref.~\cite{Sberna:2022qbn} performed full Bayesian parameter estimation for GW190521-like binaries on a circular orbit around a SMBH in LISA, also treating time-dependent Doppler and Shapiro delays. Ref.~\cite{Vijaykumar:2023tjg} extended the computation of LOSA corrections up to 3.5PN order relative to the leading LOSA term for non-spinning binaries, and Ref.~\cite{Tiwari:2025aec} further generalized them to binaries with spins aligned with the orbital angular momentum and non-zero tidal deformability. More recently, Ref.~\cite{Hendriks:2026kys} extended the LOSA modeling to a more general R{\o}mer-delay treatment that includes arbitrary outer-orbit eccentricity and time-dependent projection effects.

The first observational LOSA constraints from ground-based detectors were placed by Ref.~\cite{Vijaykumar:2023tjg} on the binary-neutron-star (BNS) events GW170817 and GW190425, yielding null results at the $|a_\parallel/c| \lesssim 10^{-6}\,\mathrm{s}^{-1}$ level using 3.5PN LOSA corrections. The same pipeline was implemented in the \texttt{Bilby\_TGR} library~\cite{Tiwari:2025aec} and used in the LVK GWTC-4.0 test of general relativity paper~\cite{LIGOScientific:2026fcf}. Ref.~\cite{Yang:2024tje} reported a claim of non-zero LOSA on GW190814 using a $4$-s analysis segment; this claim has since been refuted by Refs.~\cite{Hendriks:2026kys, Pathak:2026cik}, which showed that the evidence vanishes when a $32$-s segment is used.
These works compute LOSA corrections in the stationary-phase approximation (SPA) for the dominant $(2,2)$ mode of quasi-circular aligned-spin binaries, and either restrict the analysis to that mode or apply the same correction uniformly to all spherical-harmonic modes. For this reason, the GWTC-4.0 test of general relativity analysis adopts selection criteria on the mass ratio ($1/q \gtrsim 0.25$) and effective precessing spin ($\chi_{\rm p} \lesssim 0.4$) to restrict the LOSA inference to events where higher-mode and spin-precession effects in the underlying waveform are mild~\cite{Tiwari:2025aec, LIGOScientific:2026fcf}. Ref.~\cite{Roy:2026mco} extended this with a mode-by-mode SPA treatment and applied it to the spin-precessing model \texttt{IMRPhenomXPHM}~\cite{Pratten:2020ceb} and the inspiral-only eccentric--precessing PN model \texttt{pyEFPE}~\cite{Morras:2025nlp}, with the latter used to analyze GW200105\_162426. Orbital eccentricity has otherwise been included by Ref.~\cite{Pathak:2026cik}, which added a leading-order eccentricity phase correction to \texttt{IMRPhenomXPHM} to analyze GW190814; both eccentric analyses highlighted a degeneracy between LOSA and eccentricity.

With the exception of Refs.~\cite{Yunes:2010sm, Sberna:2022qbn}, these works apply LOSA corrections directly to frequency-domain waveform models, in the SPA and to leading order in $|\Gamma|\,t_{\rm sd}$ (with $\Gamma \equiv a_\parallel/c$ and $t_{\rm sd}$ the in-band signal duration). This treatment does not extend naturally to models natively formulated in the time domain: applying the mode-by-mode SPA corrections to a time-domain waveform would require a separate Fourier transform of each spherical-harmonic mode (21 for $\ell_{\rm max} = 4$), which is computationally expensive for long-duration signals. Furthermore, for spin-precessing sources the corrections need to be applied to the \textit{inertial-frame} modes; this forgoes the optimized routines that compute the inertial-frame polarizations directly by rotating the \textit{co-precessing-frame} modes without materializing the inertial-frame ones~\cite{Pratten:2020ceb, Ramos-buades:2023ehm}.
Time-domain models are, however, the most NR-faithful currently available when including spin precession (in the quasi-circular regime)~\cite{Varma:2019csw, Pratten:2020ceb, Estelles:2021gvs, Ramos-buades:2023ehm, Estelles:2025zah, Hamilton:2025xru} and orbital eccentricity (in the aligned-spin regime)~\cite{Gamboa:2024hli, Nagar:2024oyk, Planas:2025feq, Ramos-Buades:2026kbq, Gamboa:2026jht}, and the first IMR models combining eccentricity and precession are appearing in the time domain~\cite{Liu:2023ldr, Gamba:2024cvy, Albanesi:2025txj}. A LOSA treatment formulated natively in the time domain is therefore better suited to take advantage of these developments.

In this work, we implement LOSA corrections directly in the time domain, as a time remap of the GW polarizations: because the time remap makes no assumption about the binary dynamics, it applies uniformly to all spherical-harmonic modes and waveform morphologies, without a mode-by-mode treatment. We deploy our implementation in \texttt{SEOBNRv6EHM}~\cite{Gamboa:2026jht} for aligned-spin eccentric binaries and \texttt{SEOBNRv5PHM}~\cite{Ramos-buades:2023ehm, Estelles:2025zah} for spin-precessing quasi-circular binaries. While NR surrogate models such as \texttt{NRSur7dq4}~\cite{Varma:2019csw} are more accurate overall, their limited number of inspiral cycles makes them unsuitable for the long-duration signals from low-mass binaries most informative for LOSA inference; within this regime, \texttt{SEOBNRv6EHM} and \texttt{SEOBNRv5PHM} are the most NR-faithful models currently available on average across parameter space. The time-domain remap is model-agnostic and applies, without modification, to any other model implemented in \texttt{pySEOBNR}~\cite{Mihaylov:2023bkc}.

We validate the implementation in the quasi-circular aligned-spin (QCAS) limit by comparing against that of Refs.~\cite{Vijaykumar:2023tjg, Tiwari:2025aec} using simulated signals, and validate the model further in the quasi-circular precessing (QCP) and eccentric aligned-spin (EAS) limits while characterizing the LOSA--eccentricity and LOSA--precession degeneracies. Motivated by the potential eccentricity of the source of GW200105\_162426, we then jointly infer LOSA and orbital eccentricity for the five NSBH events in the LVK catalog through O4b~\cite{LIGOScientific:2025slb, Ligoscientific:2020zkf, Ligoscientific:2021qlt, LIGOScientific:2024elc}---GW190814, GW200105\_162426, GW200115\_042309, GW230518\_125908, and GW230529\_181500---which are promising targets for LOSA inference given their low component masses (although the nature of GW190814's secondary---NS or BH---remains uncertain~\cite{Ligoscientific:2020zkf}). To our knowledge, the LOSA constraints reported here for GW200115\_042309, GW230518\_125908, and GW230529\_181500 are the first for these events. We additionally apply the analysis to the four most significant eccentric BBH candidates identified by the \texttt{SEOBNRv6EHM} analyses of Ref.~\cite{Pompili:2026yxq}.

The remainder of this paper is organized as follows. Section~\ref{sec:methods} introduces the waveform models, describes the time-domain LOSA implementation, and outlines the Bayesian inference framework. Section~\ref{sec:validation} validates the implementation on synthetic signals from quasi-circular, eccentric, and spin-precessing binaries. Section~\ref{sec:results} presents the LOSA inference on the five NSBH events in the LVK catalog and on eccentric BBH candidates. Section~\ref{sec:conclusions} summarizes our conclusions.

\section{Methods}
\label{sec:methods}

We briefly introduce the waveform models used in this work (Sec.~\ref{sec:models}), describe the time-domain LOSA implementation (Sec.~\ref{sec:losa}), and outline the Bayesian inference framework (Sec.~\ref{sec:pe_method}).

\subsection{Waveform models}
\label{sec:models}

The waveform models employed in this work are time-domain, multipolar models built within the EOB formalism~\cite{Buonanno:1998gg, Buonanno:2000ef, Damour:2000we, Damour:2001tu, Buonanno:2005xu}, which maps the gravitational two-body problem onto the effective dynamics of a test mass in a deformed Kerr background, the deformation being  the mass ratio. The binary's conservative dynamics is governed by Hamilton's equations for the phase-space variables in the center-of-mass frame, supplemented by a radiation-reaction force describing the energy and angular-momentum losses to GW emission; the resulting inspiral is smoothly stitched to a phenomenological merger--ringdown ansatz calibrated to NR simulations.

\texttt{SEOBNRv5PHM}~\cite{Ramos-buades:2023ehm, Estelles:2025zah} is a model for BBHs in quasi-circular orbits with generic spins. It is built in the co-precessing frame upon the aligned-spin multipolar model \texttt{SEOBNRv5HM}~\cite{Pompili:2023tna}, which is calibrated to 442 quasi-circular SXS NR simulations~\cite{Scheel:2025jct}, 13 waveforms from black-hole perturbation theory, and the non-spinning second-order gravitational self-force energy flux~\cite{Vandemeent:2023ols}. The spin precession dynamics is evolved through orbit-averaged, PN-expanded spin-precession equations derived from the precessing-spin Hamiltonian~\cite{Khalil:2023kep}, and used to rotate the co-precessing-frame waveform to the inertial frame of the observer. The model includes the $(\ell, |m|) \in \{(2,2), (2,1), (3,3), (3,2), (4,4), (4,3), (5,5)\}$ spherical-harmonic modes in the co-precessing frame, including equatorial-asymmetric contributions~\cite{Estelles:2025zah}.

\texttt{SEOBNRv6EHM}~\cite{Gamboa:2026jht} is an aligned-spin model for generic planar orbits---including bound eccentric inspirals, dynamical-captures, and scattering encounters---and includes eccentricity corrections to the waveform modes and radiation-reaction force formulated in terms of the EOB phase-space variables. The eccentric orbit is parametrized by the Keplerian eccentricity $e$ and relativistic anomaly $\zeta$ at an orbit-averaged frequency. The model is calibrated only to quasi-circular SXS NR waveforms~\cite{Scheel:2025jct} and validated against 319 eccentric, one dynamical-capture, and two scattering SXS simulations, with up to an order-of-magnitude improvement in faithfulness over both \texttt{SEOBNRv5EHM} and \texttt{TEOBResumS-Dal\'i} for highly eccentric NR configurations~\cite{Gamboa:2026jht}. It includes the same set of modes as \texttt{SEOBNRv5PHM}, with the exception of $(5,5)$.

\subsection{Line-of-sight-acceleration corrections in the time domain}
\label{sec:losa}

\begin{figure}[t]
\centering
\includegraphics[width=\columnwidth]{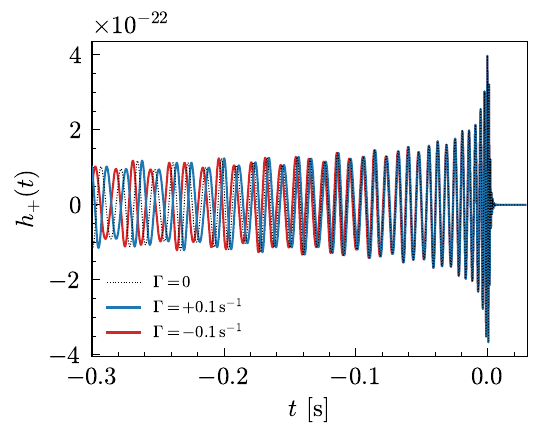}
\caption{Illustration of a constant LOSA $\Gamma\equiv a_\parallel/c$ on the plus polarization $h_+(t)$ of a time-domain waveform, aligned at the merger ($t=0$) and shown over a short time window. The reference ($\Gamma=0$) waveform is compared with $\Gamma>0$ and $\Gamma<0$, for a non-spinning binary with chirp mass $\mathcal{M}=6.5\,M_\odot$, mass ratio $q=2$, and eccentricity $e=0.30$ at $\langle f_{\rm ref} \rangle = 20\,\mathrm{Hz}$, generated with \texttt{SEOBNRv6EHM}. We take $|\Gamma|=10^{-1}\,\mathrm{s}^{-1}$, well above the prior range used in our analyses, purely to make the effect clearly visible by eye over the late inspiral. $\Gamma>0$ corresponds to the source approaching the detector (blueshift, cycles compressed) and $\Gamma<0$ to it receding (redshift, cycles stretched); the difference vanishes at merger and grows toward earlier times.}
\label{fig:td_losa}
\end{figure}

Consider a compact binary whose center of mass moves along the line of sight with velocity $v_\parallel(t)$ relative to the observer. The time at which a signal emitted at the source at time $t_{\mathrm{src}}$ arrives at the observer, $t_{\mathrm{obs}}$, differs from $t_{\mathrm{src}}$ by a R{\o}mer time of flight, $t_{\mathrm{obs}} = t_{\mathrm{src}} + D(t_{\mathrm{src}})/c$, where $D(t)$ is the source--observer separation along the line of sight~\cite{Yunes:2010sm}. For a constant line-of-sight acceleration $a_\parallel$, the velocity grows linearly with time, $v_\parallel(t) = a_\parallel t$, and the separation grows quadratically, $D(t) = D_0 + \tfrac{1}{2}\, a_\parallel t^2$. Any constant Doppler shift $v_\parallel/c$ is fully degenerate with the cosmological redshift $z_{\rm cos}$ of the source and is reabsorbed into the redshifted (detector-frame) masses~\cite{Bonvin:2016qxr, Vijaykumar:2023tjg}. Absorbing the constant offset $D_0/c$ into the origin of $t_{\mathrm{obs}}$, the relation becomes
\begin{equation}
\label{eq:t_obs}
  t_{\mathrm{obs}} = t_{\mathrm{src}} + \tfrac{1}{2}(a_\parallel/c)\, t_{\mathrm{src}}^2,
\end{equation}
whose closed-form inversion, selecting the solution with $t_{\mathrm{src}} \to t_{\mathrm{obs}}$ as $\Gamma \to 0$ (with $\Gamma \equiv a_\parallel/c$ having dimensions of inverse time), is
\begin{equation}
\label{eq:tsrc_exact}
t_{\mathrm{src}}(t_{\mathrm{obs}}) = \frac{2\, t_{\mathrm{obs}}}{\sqrt{1 + 2\,\Gamma\, t_{\mathrm{obs}}} + 1}.
\end{equation}
The observed strain is therefore a time-remapped copy of the source-frame strain,
\begin{equation}
\label{eq:td_remap}
h_{\mathrm{obs}}(t) = h_{\mathrm{src}}~\left(\frac{2\, t}{\sqrt{1 + 2\,\Gamma\, t} + 1}\right),
\end{equation}
where $t \equiv t_{\mathrm{obs}}$ and $t = 0$ marks the coalescence time. Equation~\eqref{eq:td_remap} assumes a constant-acceleration ansatz, a good approximation when a tertiary perturber is at sufficiently large separation, but does not require small $\Gamma$. Because the equation above is a pure remap of the time-domain strain, it makes no assumptions about the binary dynamics, applies uniformly to waveforms with higher-order modes, spin precession, and eccentricity, and accounts for both phase and amplitude corrections to the observed strain. Figure~\ref{fig:td_losa} illustrates the remap on an example waveform: a positive (negative) $\Gamma$ compresses (stretches) the inspiral cycles, corresponding to a source approaching (receding from) the detector.

We implement Eq.~\eqref{eq:td_remap} in the \texttt{pySEOBNR} time-domain waveform generator~\cite{Mihaylov:2023bkc} by applying the time remap to both polarizations $h_{+,\times}(t)$ using cubic spline interpolation, immediately after waveform generation, before conditioning and conversion to the frequency domain via a fast Fourier transform. Because the LOSA correction reduces to a single one-dimensional interpolation in the time domain per polarization, applied once per waveform evaluation, its cost is negligible compared to the waveform-generation cost itself.

An alternative approach suitable for EOB models is also that of Ref.~\cite{Yunes:2010sm}, which used the EOB framework to model EMRI waveforms, and inserted the Doppler factor into the EOB equations of motion during waveform generation; we instead apply the equivalent correction in post-processing. Refs.~\cite{Chamberlain:2018snj, Sberna:2022qbn} also take closely related approaches: starting from a frequency-domain model, they transform to the time domain, apply the Doppler remap there, and transform back to the frequency domain via the SPA. Ref.~\cite{Sberna:2022qbn} treats the strongly time-dependent regime beyond constant acceleration, relevant to LISA observations of stellar-mass binaries near an SMBH, and additionally includes the Shapiro time delay, i.e.\ the relativistic delay incurred by GWs traversing the SMBH's gravitational potential~\cite{Shapiro:1964uw}.

More generally, the time-remap formulation extends straightforwardly to time-dependent delays beyond the constant-acceleration ansatz, for example by including higher time-derivatives of the line-of-sight velocity. Each successive derivative contributes a modulation at a progressively lower PN order (for example, LOSA appears at $-4$PN, line-of-sight jerk at $-8$PN, line-of-sight snap at $-12$PN, and so on), which could be useful to model richer environmental scenarios~\cite{Tiwari:2024pvb, Tiwari:2025qqx}. In what follows we focus on the constant-acceleration term only, which is a good approximation when the in-band duration is much shorter than the binary's orbital period about the perturber.

Even within the constant-acceleration ansatz, the time remap neglects the effect of the source's kinematic redshift on the strain amplitude: the growing line-of-sight velocity changes the source-observer propagation distance, and the strain should be additionally rescaled by a factor $\sim (1-v_\parallel/c)$~\cite{Cusin:2024git}. The time remap in Eq.~\eqref{eq:t_obs} is a non-relativistic construction, so relativistic corrections of order $(v/c)^2$ are neglected as well. We further ignore aberration and polarization rotation from transverse source velocity~\cite{Bonvin:2022mkw,Cusin:2024git}. These effects are expected to be negligible for the accelerations and signal durations relevant to current detectors.

\subsection{Bayesian inference framework}
\label{sec:pe_method}

We follow the standard Bayesian GW parameter-estimation framework adopted in recent LVK analyses~\cite{LIGOScientific:2026ifv}, and the conventions of Ref.~\cite{Pompili:2026yxq} for aspects specific to eccentric sources.
For aligned-spin eccentric binaries, \texttt{SEOBNRv6EHM} is parametrized by $\{m_1, m_2, \chi_1, \chi_2, e, \zeta\}$, with the Keplerian eccentricity $e$ and relativistic anomaly $\zeta$~\cite{Darwin1959gravity} specified at an orbit-averaged reference frequency $\langle f_{\rm{ref}}\rangle$~\cite{Ramos-Buades:2023yhy}.
Throughout this work, we report results in terms of the derived mass and spin parameters: the total mass $M = m_1 + m_2$, chirp mass $\mathcal{M} = (m_1 m_2)^{3/5}/(m_1 + m_2)^{1/5}$, mass ratio $q = m_1/m_2 \geq 1$, effective spin $\chi_{\rm{eff}} = (m_1 \chi_1 + m_2 \chi_2)/(m_1 + m_2)$, and the effective precessing spin $\chi_{\rm p}$~\cite{Schmidt:2014iyl}. Unless stated otherwise, we quote detector-frame (redshifted) masses.
We adopt uniform priors on $e$ and $\zeta$, as well as a uniform prior $\Gamma \in [-10^{-2}, +10^{-2}]\,\mathrm{s}^{-1}$. The remaining priors follow standard LVK conventions~\cite{LIGOScientific:2026ifv}.

All inference is performed with \texttt{Bilby}~\cite{Ashton:2018jfp, Romero-shaw:2020owr} and the \texttt{dynesty} sampler~\cite{Speagle:2019ivv}, as well as \texttt{parallel Bilby}~\cite{Smith:2019ucc} for some of the longer NSBH signals. We adopt the \texttt{acceptance-walk} stepping method with $\texttt{naccept}=60$ and $\texttt{nlive}=1000$, apply distance marginalization, and otherwise follow LVK analyses~\cite{LIGOScientific:2026ifv}.

\section{Validation on synthetic signals}
\label{sec:validation}

We validate the time-domain LOSA implementation through analyses of simulated signals with zero and non-zero LOSA, including QCAS, QCP, and EAS binaries. In all recoveries throughout this section, the LOSA parameter $\Gamma$ is included as a free parameter. The injections are performed in zero noise for the LIGO Hanford and Livingston detectors, assuming the A+ power spectral density (PSD)~\cite{Ligo:noisecurves}, corresponding to the fifth observing run (O5) of the LVK network.

All injections share a common base configuration: a low-mass BBH with chirp mass $\mathcal{M} = 6.5\,M_\odot$ and mass ratio $q = 2$ (component masses $m_1 \approx 10.4\,M_\odot$, $m_2 \approx 5.2\,M_\odot$), at luminosity distance $d_L = 600\,\mathrm{Mpc}$, yielding a $32$-s signal from $f_{\rm start} = 20\,\mathrm{Hz}$ with network signal-to-noise ratio (SNR) $\rho \approx 40$. The low chirp mass and long signal duration allow $\Gamma$ to be constrained within the prior range at this sensitivity. The QCAS injection has $\chi_1 = \chi_2 = 0$ and $e = 0$; the EAS injection extends the QCAS one with $e_{\rm gw} = 0.3$ at $\langle f_{\rm ref}\rangle = 20\,\mathrm{Hz}$; and the QCP injection extends it with dimensionless spin magnitudes $\chi_{1,2} = 0.8$ tilted at $\theta_{1,2} = \pi/3$ relative to the orbital angular momentum (effective precessing spin $\chi_{\rm p} \approx 0.7$). Each configuration is injected both with $\Gamma = 0$ and with $\Gamma = 10^{-3}\,\mathrm{s}^{-1}$.

\subsection{Quasi-circular aligned-spin injections}
\label{sec:val_qcas}

\begin{figure}[t]
\centering
\includegraphics[width=1.0\linewidth]{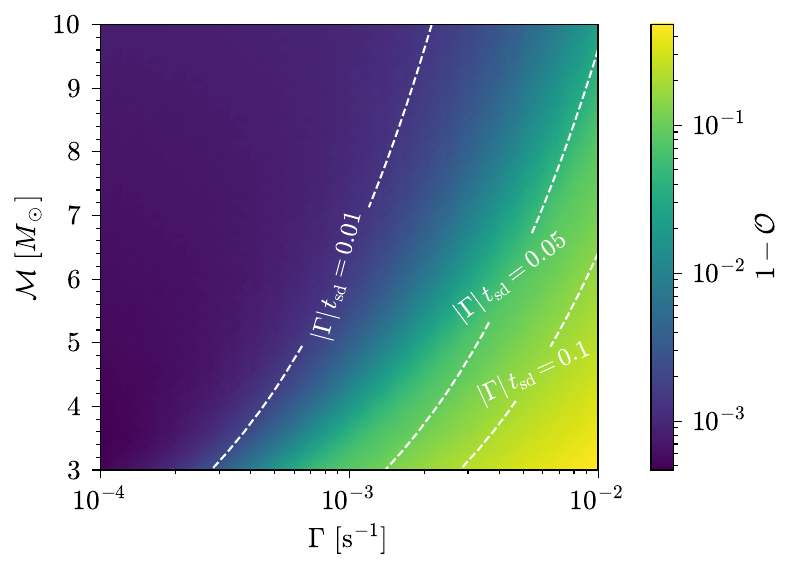}
\caption{Mismatch $1 - \mathcal{O}$ (with $\mathcal{O}$ the noise-weighted overlap between two waveforms, maximized over time and phase) between the time-domain LOSA implementation in \texttt{SEOBNRv6EHM} and the frequency-domain one in \texttt{Bilby\_TGR}~\cite{Vijaykumar:2023tjg, Tiwari:2025aec} applied to \texttt{IMRPhenomXAS}, as a function of the LOSA parameter $\Gamma$ and detector-frame chirp mass $\mathcal{M}$. Overlaps are computed against the A+ design PSD for non-spinning quasi-circular binaries with $q = 2$ and the dominant $(2,2)$ mode. White dashed contours mark $|\Gamma|\,t_{\rm sd}$, with $t_{\rm sd}$ the signal duration from $f_{\rm start} = 20\,\mathrm{Hz}$. At small $|\Gamma|\,t_{\rm sd}$, the mismatch approaches the floor set by the underlying difference between \texttt{SEOBNRv6EHM} and \texttt{IMRPhenomXAS} in the non-accelerating limit; the two LOSA implementations agree to near this level throughout $|\Gamma|\,t_{\rm sd} \lesssim 0.01$ (the validity range of the leading-order treatment assumed in \texttt{Bilby\_TGR}) and depart smoothly above this.}
\label{fig:mismatch}
\end{figure}

As a first validation, we compare our time-domain LOSA implementation against the frequency-domain one in \texttt{Bilby\_TGR}~\cite{Vijaykumar:2023tjg, Tiwari:2025aec} at the waveform level, by computing mismatches (Fig.~\ref{fig:mismatch}). The two agree closely throughout the region $|\Gamma|\,t_{\rm sd} \lesssim 0.01$, where $t_{\rm sd}$ is the signal duration from $f_{\rm start} = 20\,\mathrm{Hz}$, and depart smoothly above this; \texttt{Bilby\_TGR} adopts a leading-order approximation in $|\Gamma|\,t_{\rm sd}$, while our implementation makes no such assumption. Consistent with this, prior LOSA analyses using the \texttt{Bilby\_TGR} pipeline adopt $|\Gamma|\,t_{\rm sd} \lesssim 0.01$ as a selection criterion for the events analyzed~\cite{Tiwari:2025aec, LIGOScientific:2026fcf}.

\begin{figure}[t]
\centering
\includegraphics[width=0.95 \linewidth]{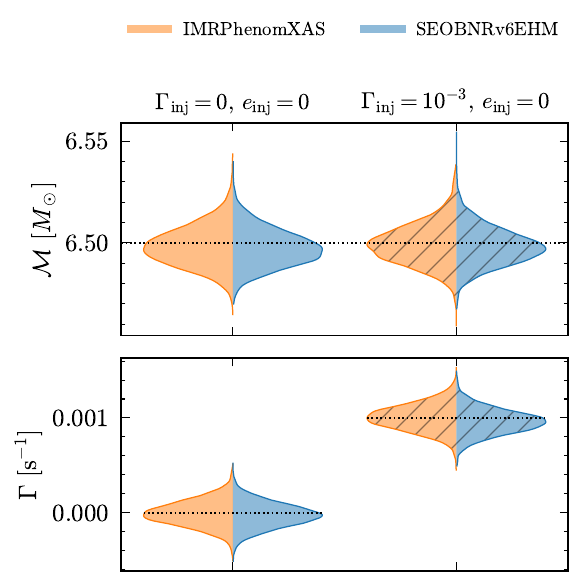}
\caption{Validation of the time-domain LOSA implementation in \texttt{SEOBNRv6EHM} against the frequency-domain implementation in \texttt{Bilby\_TGR}~\cite{Vijaykumar:2023tjg, Tiwari:2025aec}.
	A signal from a non-spinning, quasi-circular binary generated with \texttt{IMRPhenomXAS}, with both $\Gamma=0$ and $\Gamma=10^{-3}\,\mathrm{s}^{-1}$ applied through the prescription of Refs.~\cite{Vijaykumar:2023tjg, Tiwari:2025aec}, is injected in zero noise and recovered with both pipelines.
	Split violins show the recovered marginal posteriors on the chirp mass $\mathcal{M}$ (top) and the LOSA parameter $\Gamma$ (bottom), with the left and right columns corresponding to the $\Gamma_{\rm inj} = 0$ and $\Gamma_{\rm inj} = 10^{-3}\,\mathrm{s}^{-1}$ (hatched) injections; dotted lines mark the injected values.
	The recoveries are consistent in both the zero- and non-zero-$\Gamma$ cases.
	}
	\label{fig:val_qcas}
\end{figure}

We then inject a signal from a non-spinning, quasi-circular binary generated with the QCAS model \texttt{IMRPhenomXAS}~\cite{Pratten:2020fqn}, with both $\Gamma = 0$ and $\Gamma = 10^{-3}\,\mathrm{s}^{-1}$ applied through the LOSA prescription implemented in \texttt{Bilby\_TGR}~\cite{Vijaykumar:2023tjg, Tiwari:2025aec}. For the injected $\Gamma = 10^{-3}\,\mathrm{s}^{-1}$, $|\Gamma|\,t_{\rm sd} \approx 0.01$, near the boundary in Fig.~\ref{fig:mismatch} above which the two implementations start to depart.
We recover the injection with the same model and, separately, with \texttt{SEOBNRv6EHM} restricted to the $e = 0$ limit and to the dominant $(\ell, m) = (2,2)$ mode, in both cases allowing aligned spins in the prior. This comparison serves as additional validation of the time-domain LOSA implementation against the established frequency-domain pipeline, in the limit where the two should agree.

The recovered posteriors on $\Gamma$ and on the chirp mass are shown in Fig.~\ref{fig:val_qcas}; the two pipelines are in excellent agreement in both the zero- and non-zero-$\Gamma$ cases, with all other source parameters recovered consistently as well.

\begin{figure*}[t]
\centering
\subfloat[Quasi-circular precessing injection]{\includegraphics[width=0.45\linewidth]{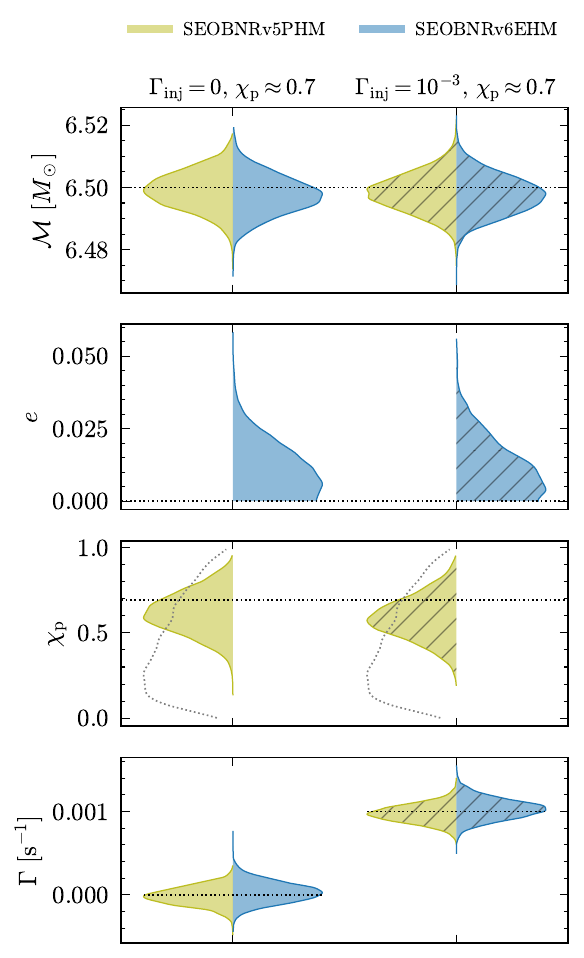}\label{fig:val_qcp}}\hfill
\subfloat[Eccentric aligned-spin injection]{\includegraphics[width=0.45\linewidth]{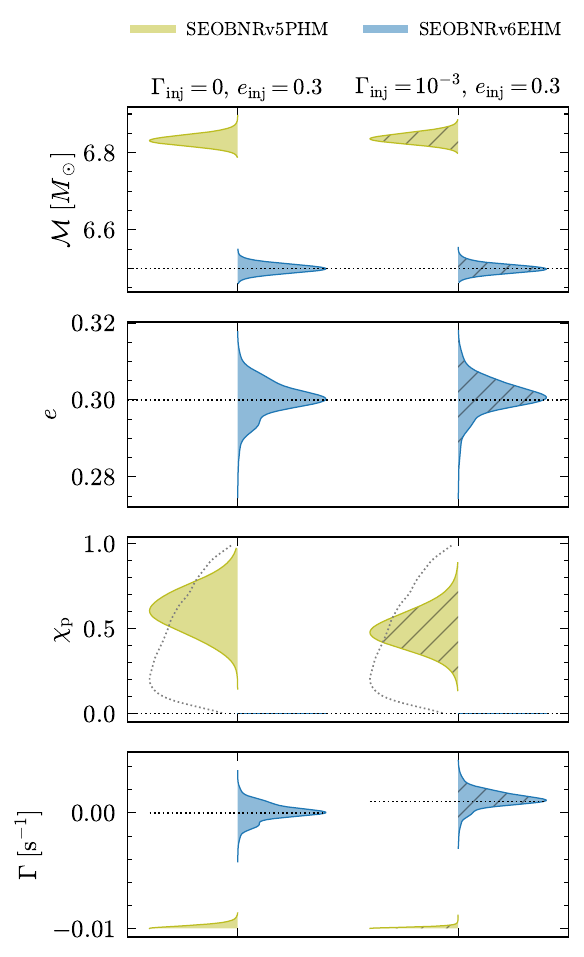}\label{fig:val_eas}}
\caption{Validation on the quasi-circular precessing (QCP) (left) and eccentric aligned-spin (EAS) (right) injections, with recoveries under \texttt{SEOBNRv5PHM} (yellow) and \texttt{SEOBNRv6EHM} (blue); both recovery templates include the LOSA parameter $\Gamma$.
  The QCP signal ($\chi_{1,2} = 0.8$, $\theta_{1,2} = \pi/3$, $\chi_{\rm p} \approx 0.7$) is generated with \texttt{SEOBNRv5PHM}; the EAS signal ($e = 0.3$ at $\langle f_{\rm ref}\rangle = 20\,\mathrm{Hz}$, non-spinning) with \texttt{SEOBNRv6EHM}. Each is injected in zero noise for $\Gamma = 0$ (left column) and $\Gamma = 10^{-3}\,\mathrm{s}^{-1}$ (right column, hatched).
  Split violins show the marginal posteriors on $\mathcal{M}$, $e$ (only for \texttt{SEOBNRv6EHM}), $\chi_{\rm p}$ (only for \texttt{SEOBNRv5PHM}), and $\Gamma$; the gray dotted curve overlaid on the $\chi_{\rm p}$ panel is the prior, and horizontal dotted lines mark the injected values.
  }
\end{figure*}

\subsection{Quasi-circular precessing injections}
\label{sec:val_qcp}

We inject a signal generated with \texttt{SEOBNRv5PHM} from the QCP binary configuration described above ($\chi_{1,2} = 0.8$, $\theta_{1,2} = \pi/3$, $\chi_{\rm p} \approx 0.7$), with both $\Gamma = 0$ and $\Gamma = 10^{-3}\,\mathrm{s}^{-1}$. We recover the injection with two templates: \texttt{SEOBNRv5PHM} itself, to validate the time-domain LOSA implementation in the spin-precessing regime with higher-order modes; and the EAS model \texttt{SEOBNRv6EHM}, which probes whether spin precession can be misattributed to orbital eccentricity or to a spurious LOSA when the recovery template lacks precession.

The recovered posteriors are shown in Fig.~\ref{fig:val_qcp}. The \texttt{SEOBNRv5PHM} recovery yields posteriors consistent with the injected values for $\Gamma$ and all other source parameters in both the zero- and non-zero-$\Gamma$ cases, although the $\chi_{\rm p}$ posterior does not peak at the injected value, likely a consequence of the non-uniform $\chi_{\rm p}$ prior (gray dotted curve). The \texttt{SEOBNRv6EHM} recovery is shown alongside: in both $\Gamma$ cases the chirp mass is unbiased, the LOSA posterior is centered on the injected value with width comparable to the \texttt{SEOBNRv5PHM} recovery, and the inferred eccentricity is consistent with zero. Spin precession is therefore neither misattributed to a spurious LOSA nor absorbed into a spurious eccentricity. In a geometric picture~\cite{Cutler:2007mi}, the waveform difference between the spin-precessing injection and the EAS template decomposes into a component tangent to the template manifold, which shifts the recovered parameters, and an orthogonal component, which only lowers the likelihood and SNR: here the difference is predominantly orthogonal, so it appears as a modest loss of recovered network SNR ($\Delta \rho \approx 0.5$) rather than as a parameter bias.

\subsection{Eccentric aligned-spin injections}
\label{sec:val_eas}

We finally inject a signal generated with \texttt{SEOBNRv6EHM} from the EAS binary configuration described above ($e = 0.3$ at $\langle f_{\rm ref}\rangle = 20\,\mathrm{Hz}$, non-spinning), with both $\Gamma = 0$ and $\Gamma = 10^{-3}\,\mathrm{s}^{-1}$. We recover the injection with two templates: \texttt{SEOBNRv6EHM} itself, as a validation of the LOSA implementation in the EAS regime; and the QCP model \texttt{SEOBNRv5PHM}, to probe whether orbital eccentricity can be misattributed to spin precession or to a spurious LOSA when the recovery template lacks eccentricity.

The recovered marginal posteriors are shown in Fig.~\ref{fig:val_eas}.
The \texttt{SEOBNRv6EHM} recovery yields posteriors consistent with the injected values for $\mathcal{M}$, $e$, and the remaining source parameters in both the zero- and non-zero-$\Gamma$ cases. The $\Gamma$ posterior, however, is broadened by a factor of $\sim 6$ relative to the recovery of the QCP injected-signal (see bottom left panel): the 90\% credible width goes from $\sim 0.4 \times 10^{-3}\,\mathrm{s}^{-1}$ to $\sim 2.4 \times 10^{-3}\,\mathrm{s}^{-1}$. This reflects a degeneracy between $\Gamma$ and $e$ in the inspiral phase~\cite{Pathak:2026cik, Roy:2026mco}, which inflates the joint posterior along the degeneracy direction when $e$ is non-trivially measured; a subdominant contribution comes from the shorter in-band inspiral of an eccentric signal, which carries fewer cycles and hence less information on $\Gamma$. The eccentricity is broadened by the same degeneracy, although the effect would be apparent only relative to a recovery without LOSA; a non-zero eccentricity is generically measured more tightly than a vanishing one~\cite{Pompili:2026yxq}. As a consequence, the non-zero injection $\Gamma = 10^{-3}\,\mathrm{s}^{-1}$ is recovered with the posterior peaked near the injected value, $\Gamma = 1.1^{+1.2}_{-1.2} \times 10^{-3}\,\mathrm{s}^{-1}$, but with a 90\% credible interval that includes zero: at this SNR ($\rho \approx 40$) the LOSA--eccentricity degeneracy prevents a confident separation of $\Gamma = 10^{-3}\,\mathrm{s}^{-1}$ from zero in an eccentric source.

The \texttt{SEOBNRv5PHM} recovery is qualitatively different from the QCP injected-signal case: the absence of eccentricity in the template introduces substantial biases across the parameter space. The chirp mass is overestimated by $\Delta\mathcal{M} \sim 0.3\,M_\odot$, with the injected value well outside the 90\% posterior support, consistent with the known $\mathcal{M}$--$e$ degeneracy~\cite{Favata:2021vhw}. The effective precessing spin is spuriously inferred from this non-spinning source: although the non-uniform $\chi_{\rm p}$ prior (gray dotted curve) already pushes the posterior toward intermediate values, the recovered $\chi_{\rm p}$ peaks at $\sim 0.5$--$0.6$, above the mode of the prior itself. A spurious aligned-spin component is also inferred ($\chi_{\rm eff} \sim 0.15$). Finally, the LOSA parameter is driven to the lower prior boundary $\Gamma = -10^{-2}\,\mathrm{s}^{-1}$, independently of whether the injected $\Gamma$ was zero or non-zero. Eccentricity thus acts as a LOSA mimicker, with the negative direction of the bias consistent with analogous BNS injection results of Ref.~\cite{Tiwari:2025aec} (their Fig.~5). Conversely, LOSA can mimic a spurious eccentricity~\cite{Bhat:2025lri}; this degeneracy can be broken by tracking the eccentricity evolution with frequency~\cite{Bhat:2025lri}, but is most directly addressed by the joint inference of both effects in a waveform model that includes them, as in our \texttt{SEOBNRv6EHM} analyses.

Taken together, the QCP and EAS injection studies quantify the asymmetry between the precession-into-eccentricity and eccentricity-into-precession degeneracies, which is relevant for LOSA inference using either kind of template alone. In the absence of a model that includes both effects jointly, the EAS template seems a more robust choice, since precession-induced biases on the recovered LOSA are markedly milder than the eccentricity-induced ones. A more realistic test would inject a signal carrying \emph{both} precession and eccentricity simultaneously; we leave this to future work with the forthcoming eccentric--precessing \texttt{SEOBNRv6EPHM} model.

\section{Analysis of real events}
\label{sec:results}

We now apply our model to selected events from the LVK catalog, focusing on analyses that jointly infer eccentricity and LOSA with \texttt{SEOBNRv6EHM}. As shown by the \texttt{SEOBNRv6EHM} recovery of the strongly-precessing injection considered in Sec.~\ref{sec:val_qcp}, for long-duration signals of the kind considered here, spin precession is not misattributed to a spurious LOSA, and an injected LOSA is correctly recovered by the EAS template. Conversely, the \texttt{SEOBNRv5PHM} recovery of the eccentric injection in Sec.~\ref{sec:val_eas} shows that the absence of eccentricity in the recovery template can lead to a spurious LOSA and to biases in other parameters. We therefore adopt \texttt{SEOBNRv6EHM} as the primary template for LOSA inference, and additionally show for the eccentric NSBH candidate GW200105\_162426 that a \texttt{SEOBNRv5PHM} analysis yields spurious support for non-zero $\Gamma$, in agreement with the injection study.

For each event, we use the strain data, calibration uncertainties, and PSDs provided by the Gravitational Wave Open Science Center (GWOSC)~\cite{LIGOScientific:2019lzm}. For events with data quality issues, we use deglitched frames provided in the respective data releases~\cite{ligo_scientific_collaboration_and_virgo_2022_6477076, ligo_scientific_collaboration_and_virgo_2021_5546680, ligo_scientific_collaboration_2025_16857060}. The sampling rate, segment length, and frequency settings are adapted to each event and closely follow the LVK analyses.

The bulk of the section is devoted to the five NSBH events through O4b: GW200105\_162426 and GW200115\_042309~\cite{Ligoscientific:2021qlt}, GW190814~\cite{Ligoscientific:2020zkf}, GW230518\_125908~\cite{LIGOScientific:2025slb}, and GW230529\_181500~\cite{LIGOScientific:2024elc}.
We analyze the events with the BBH model \texttt{SEOBNRv6EHM}, neglecting any tidal deformability of the NS component. Tidal corrections enter the GW phase at $5$PN~\cite{Flanagan:2007ix}, and are not strongly correlated with the LOSA contribution entering at $-4$PN. For a simulated BNS signal, Ref.~\cite{Tiwari:2025aec} found that neglecting tides biases $\Gamma$ by $\sim 10^{-6}\,\mathrm{s}^{-1}$ at network SNR $\rho \sim 50$. For our NSBH sample the effective tidal parameter is suppressed relative to a BNS, and the bias on $\Gamma$ is expected to be several orders of magnitude smaller than the statistical uncertainties reported below.

We additionally analyze the four most significant eccentric BBH candidates identified by Ref.~\cite{Pompili:2026yxq} (GW200129\_065458, GW200208\_222617, GW190701\_203306, and GW231223\_032836) under the prior $\Gamma \in [-10^{-2}, +10^{-2}]\,\mathrm{s}^{-1}$. All four yield uninformative LOSA posteriors, with $\log_{10}\mathcal{B}^{\Gamma\neq 0}_{\Gamma=0}$ compatible with zero within the nested-sampling evidence uncertainty, as expected for the higher masses and shorter signal durations of these binaries relative to the NSBH sample~\cite{Bonvin:2016qxr}; we therefore do not report the LOSA posteriors in detail. The inferred eccentricities are consistent with the analyses of Ref.~\cite{Pompili:2026yxq} without LOSA, with $e = 0.26^{+0.05}_{-0.07}$, $0.35^{+0.11}_{-0.14}$, $0.53^{+0.08}_{-0.34}$, and $0.47^{+0.07}_{-0.24}$ at $\langle f_{\rm ref}\rangle = 10\,\mathrm{Hz}$ (median and 90\% credible interval), respectively: the eccentricity measurements for these candidates are robust to allowing a non-zero $\Gamma$.

\subsection{NSBH candidates through O4b}
\label{sec:res_NSBH}

\begin{figure}[t]
\centering
\includegraphics[width=0.9 \linewidth]{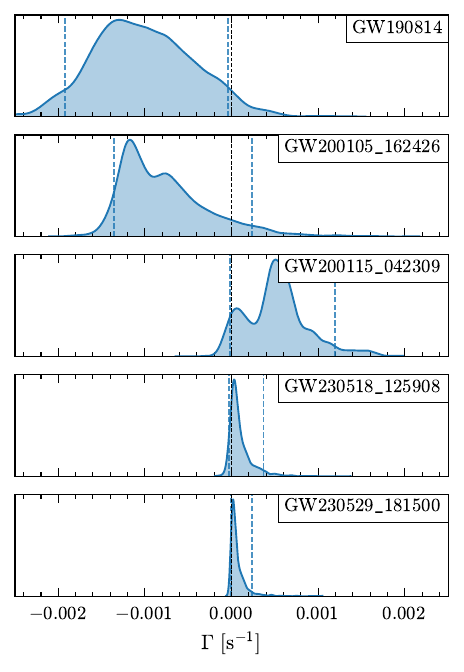}
\caption{
  Marginal posteriors on the LOSA parameter $\Gamma$ for the five NSBH events in the LVK catalog through O4b, from the analysis with \texttt{SEOBNRv6EHM}. Dashed blue verticals mark the 90\% credible intervals; the dashed black vertical marks $\Gamma = 0$.
}
\label{fig:res_NSBH}
\end{figure}

The marginal $\Gamma$ posteriors for the full NSBH sample are summarized in Fig.~\ref{fig:res_NSBH}, and a summary of the inferred parameters and Bayes factors is collected in Table~\ref{tab_nsbh}. All five events yield $\Gamma$ posteriors consistent with zero at the 90\% credibility, except for GW190814, where $\Gamma = 0$ lies just outside the 90\% credible interval. We also compute the Bayes factor between the $\Gamma\neq 0$ and $\Gamma = 0$ hypotheses via the Savage--Dickey density ratio~\cite{Sd_ratio} at $\Gamma = 0$ (Table~\ref{tab_nsbh}); in all five cases the $\Gamma = 0$ hypothesis is favored, with $\log_{10}\mathcal{B}^{\Gamma\neq 0}_{\Gamma=0}$ ranging from $-0.56$ for GW190814 to $-2.22$ for GW230529\_181500.

The width of the $\Gamma$ constraint scales with the signal duration: GW230529\_181500 and GW230518\_125908, analyzed using $128$-s segments, are constrained at $|\Gamma| \lesssim 3 \times 10^{-4}\,\mathrm{s}^{-1}$ (90\% credibility), while GW190814, analyzed over a $32$-s segment, yields the broadest bound, $|\Gamma| \lesssim 2 \times 10^{-3}\,\mathrm{s}^{-1}$.
By comparison, the BNS analyses of Ref.~\cite{Vijaykumar:2023tjg} on GW170817 and GW190425 reach $|\Gamma| \lesssim 10^{-6}\,\mathrm{s}^{-1}$, two to three orders of magnitude tighter than the NSBH bounds reported here, reflecting the longer signal durations and lower chirp masses of these sources.
The Bayes factor in favor of $\Gamma = 0$ follows the same hierarchy: tighter $\Gamma$ posteriors imply a larger Occam penalty against the $\Gamma\neq 0$ hypothesis. The inferred eccentricities are small in all cases, with medians at $e \lesssim 0.06$.

GW200115\_042309 was also examined by the LVK GWTC-4.0 test of GR analysis~\cite{LIGOScientific:2026fcf}, which reported it as consistent with zero LOSA at slightly less than 90\% credibility but did not quote a quantitative constraint, applying a cutoff $|\Gamma|\,t_{\rm sd} \lesssim 0.01$ to enforce validity of the leading-order expansion in $\Gamma$ adopted there. Our implementation does not assume $|\Gamma|\,t_{\rm sd} \ll 1$; we therefore quote the GW200115\_042309 constraint despite $|\Gamma|\,t_{\rm sd} \lesssim 0.05$ at the 90\% bound. In the remainder of this section we examine GW200105\_162426 and GW190814 in more detail, and compare with previous analyses of these events.

\renewcommand{\arraystretch}{1.6}
\begin{table}[t]
\centering
\small
\begin{tabular*}{\columnwidth}{@{\extracolsep{\fill}} l c c c c}
\hline\hline
Event & $\mathcal{M}\ [M_\odot]$ & $\Gamma\ [10^{-3}\,\mathrm{s}^{-1}]$ & $e$ & $\log_{10}\mathcal{B}^{\Gamma\neq 0}_{\Gamma=0}$ \\
\hline
GW190814         & $6.38^{+0.05}_{-0.04}$ & $-1.04^{+1.00}_{-0.88}$ & $0.03^{+0.05}_{-0.03}$ & $-0.56^{+0.04}_{-0.04}$ \\
GW200105\_162426 & $3.59^{+0.02}_{-0.02}$ & $-0.88^{+1.11}_{-0.48}$ & $0.06^{+0.07}_{-0.06}$ & $-0.60^{+0.03}_{-0.03}$ \\
GW200115\_042309 & $2.59^{+0.01}_{-0.02}$ & $0.51^{+0.69}_{-0.53}$ & $0.06^{+0.08}_{-0.05}$ & $-1.14^{+0.04}_{-0.05}$ \\
GW230518\_125908 & $2.94^{+0.01}_{-0.01}$ & $0.05^{+0.31}_{-0.08}$ & $0.03^{+0.06}_{-0.03}$ & $-1.98^{+0.02}_{-0.03}$ \\
GW230529\_181500 & $2.027^{+0.005}_{-0.004}$ & $0.04^{+0.20}_{-0.04}$ & $0.03^{+0.06}_{-0.03}$ & $-2.22^{+0.06}_{-0.06}$ \\
\hline\hline
\end{tabular*}
\caption{Inferred parameters and Bayes factor for the five NSBH events in the LVK catalog through O4b, from the analysis with \texttt{SEOBNRv6EHM}. The detector-frame chirp mass $\mathcal{M}$, the LOSA parameter $\Gamma$, and the orbital eccentricity $e$ at $\langle f_{\rm ref} \rangle = 20\,\mathrm{Hz}$ are quoted as the posterior median with symmetric 90\% credible interval. The Bayes factor between the $\Gamma\neq 0$ and $\Gamma=0$ hypotheses is computed from the Savage--Dickey density ratio at $\Gamma = 0$, with the quoted uncertainty estimated by bootstrap resampling of the posterior.}
\label{tab_nsbh}
\end{table}

\begin{figure*}[t]
\centering
\subfloat[GW200105\_162426]{\includegraphics[width=0.48\linewidth]{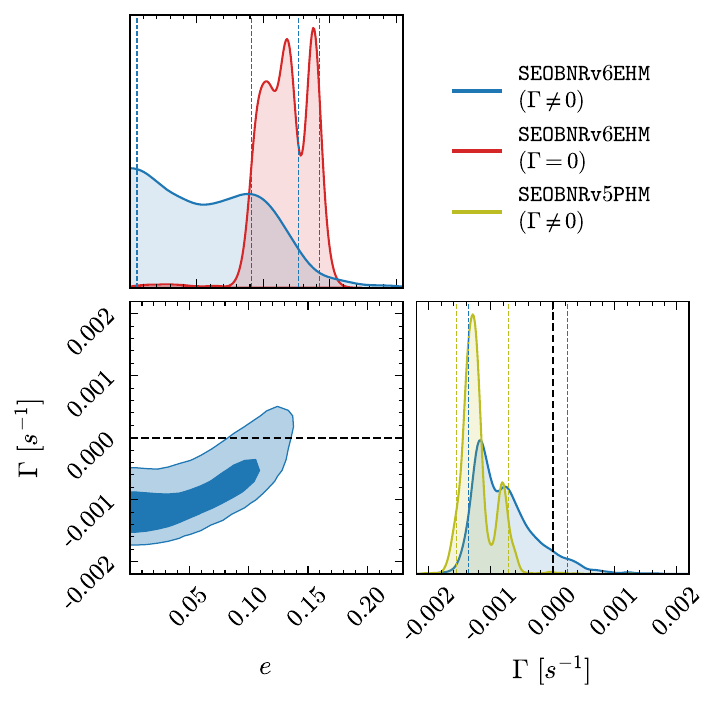}\label{fig:res_corners_200105}}
\hfill
\subfloat[GW190814]{\includegraphics[width=0.48\linewidth]{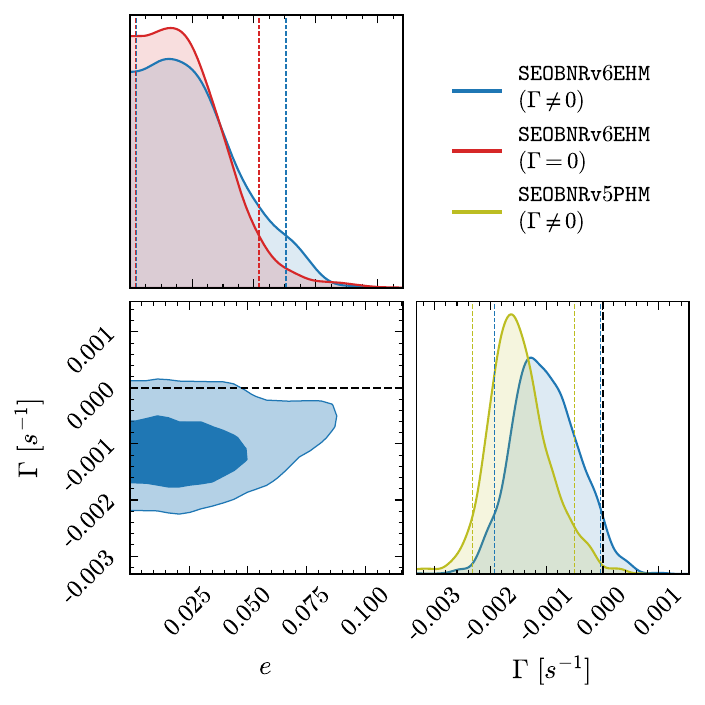}\label{fig:res_corners_190814}}
\caption{
  Joint inference of the LOSA parameter $\Gamma$ and the orbital eccentricity $e$ at $\langle f_{\rm ref} \rangle = 20\,\mathrm{Hz}$ for GW200105\_162426 (left) and GW190814 (right). Blue contours show 50\% and 90\% credible regions of the two-dimensional $(e, \Gamma)$ posterior from \texttt{SEOBNRv6EHM}, with one-dimensional marginals on the side panels. Overlaid are the marginal on $e$ from a \texttt{SEOBNRv6EHM} analysis with $\Gamma$ fixed to zero (red), and the marginal on $\Gamma$ from a complementary \texttt{SEOBNRv5PHM} analysis (yellow). Dashed colored vertical lines mark the 90\% credible intervals; the dashed black line marks $\Gamma = 0$. For GW200105\_162426, the joint posterior is incompatible with simultaneous $\Gamma = 0$ and $e = 0$ at 90\% credibility; the same point lies within the 90\% credible region for GW190814.
}
\end{figure*}

\subsubsection{Analysis of GW200105\_162426}

For GW200105\_162426, multiple independent analyses have indicated signs of non-zero orbital eccentricity, with some sensitivity to prior choices~\cite{Morras:2025xfu, Planas:2025plq, Jan:2025fps, Kacanja:2025kpr, Pompili:2026yxq, Clarke:2026cuw}.
Ref.~\cite{Roy:2026mco} additionally reported a $\Gamma$ constraint with the PN inspiral-only eccentric--precessing model \texttt{pyEFPE}, finding a broadening of the eccentricity posterior when $\Gamma$ is allowed to vary, and support for $e \to 0$ in the non-zero-$\Gamma$ region of the joint posterior.
Our analysis provides the first LOSA inference for GW200105\_162426 with an eccentric IMR multipolar template. The resulting joint and marginal posteriors are shown in Fig.~\ref{fig:res_corners_200105}: the $(\Gamma, e)$ joint posterior displays a correlation between LOSA and eccentricity, with $\Gamma = 0$ lying inside the 90\% credible region at non-zero $e$, and, conversely, $e = 0$ lying inside it at non-zero $\Gamma$; while both being zero simultaneously is disfavored at the 90\% level. The constraint $\Gamma = -0.88^{+1.11}_{-0.48} \times 10^{-3}\,\mathrm{s}^{-1}$ is broadly consistent with that of Ref.~\cite{Roy:2026mco}; some difference is expected, since \texttt{SEOBNRv6EHM} and other IMR multipolar eccentric models already yield $e$ posteriors that slightly differ from \texttt{pyEFPE} in the absence of LOSA~\cite{Morras:2025xfu, Planas:2025plq, Jan:2025fps, Kacanja:2025kpr, Pompili:2026yxq, Clarke:2026cuw}. The eccentricity-only analysis (with $\Gamma$ fixed to zero; red curve in Fig.~\ref{fig:res_corners_200105}) recovers a multimodal posterior on $e$, consistent among analyses using IMR multipolar models.

The corresponding Bayes factor (Table~\ref{tab_nsbh}) is $\log_{10}\mathcal{B}^{\Gamma\neq 0}_{\Gamma=0} \approx -0.60$: the eccentric, zero-LOSA hypothesis is preferred over its $\Gamma \neq 0$ extension. In combination with the eccentricity Bayes factors reported by Ref.~\cite{Pompili:2026yxq}---$\log_{10}\mathcal{B}^{e\neq 0}_{e=0} \approx 0.85$ relative to the QCAS baseline and $\approx 0.67$ relative to the QCP baseline---the eccentric, non-accelerating hypothesis is preferred over both the eccentric, accelerating and the quasi-circular alternatives. A complementary \texttt{SEOBNRv5PHM} analysis recovers a $\Gamma$ posterior peaked away from zero (Fig.~\ref{fig:res_corners_200105}), with $\log_{10}\mathcal{B}^{\Gamma\neq 0}_{\Gamma=0} \approx 0.45$---a weak but positive preference for non-zero LOSA, qualitatively consistent with the eccentricity-induced LOSA bias seen on the EAS injection (Sec.~\ref{sec:val_eas}). The Bayes factor between the two $\Gamma \neq 0$ analyses, computed directly from the nested-sampling evidences, is $\log_{10}\mathcal{B}^{\texttt{v6EHM},\,\Gamma\neq 0}_{\texttt{v5PHM},\,\Gamma\neq 0} \approx 0.34$. Combined with the \texttt{SEOBNRv6EHM} value above, this yields
\begin{equation*}
\log_{10}\mathcal{B}_{\texttt{v6EHM},\,\Gamma = 0}^{\texttt{v5PHM},\,\Gamma\neq 0} \approx - 0.60 - 0.34 \approx - 0.94,
\end{equation*}
so the eccentric, non-accelerating hypothesis remains the preferred interpretation.
We caution that the eccentricity inference for this event is sensitive to prior choices~\cite{Morras:2025xfu, Planas:2025jny, Clarke:2026cuw}. A uniform prior reflects the likelihood support in the data, whereas a log-uniform prior concentrates probability at small eccentricities and, for a signal with marginal eccentricity support such as this one, weakens the preference for eccentricity, with a residual dependence on its lower bound and on the reference frequency at which eccentricity is defined. A uniform $\Gamma$ prior similarly provides a LOSA constraint driven by the data rather than the prior, and a prior weighted toward the small $\Gamma$ values predicted astrophysically is expected to similarly reduce the support for a non-zero LOSA.

\subsubsection{Analysis of GW190814}

GW190814 has been analyzed several times for LOSA in the recent literature, with conflicting conclusions whose differences trace to choices of signal duration and waveform modeling systematics; we briefly summarize them before placing our result. Among the five events, GW190814 is the only one with a previous LOSA claim in the literature: Ref.~\cite{Yang:2024tje} reported $\log_{10}\mathcal{B}^{\Gamma\neq 0}_{\Gamma=0} \approx 1.8$ in favor of a non-zero LOSA using a $4$-s analysis segment, subsequently refuted by Refs.~\cite{Pathak:2026cik, Hendriks:2026kys} when a $32$-s segment is used (with Ref.~\cite{Pathak:2026cik} additionally including a leading-order eccentricity phase correction in the recovery waveform). More recently, Ref.~\cite{Roy:2026mco} revisited the analysis with the same $32$-s segment but applied LOSA correction mode-by-mode to \texttt{IMRPhenomXPHM}, rather than via the quadrupolar prescription applied uniformly to all modes~\cite{Yang:2024tje, Pathak:2026cik, Hendriks:2026kys}, identifying this as a source of systematic bias. In their mode-consistent \texttt{IMRPhenomXPHM} analysis, $\Gamma = 0$ lies slightly outside the 90\% credible interval, in mild tension with the analyses of Refs.~\cite{Pathak:2026cik, Hendriks:2026kys}; the corresponding logarithmic Bayes factor in favor of LOSA nonetheless remains below unity, and Ref.~\cite{Roy:2026mco} concludes that the evidence for non-zero LOSA is not statistically significant.

Here, we revisit the analysis with an eccentric IMR multipolar template, using the same $32$-s segment as in the LVK analysis~\cite{Ligoscientific:2020zkf}. The joint $(e, \Gamma)$ posterior is shown in Fig.~\ref{fig:res_corners_190814}: the inferred eccentricity is consistent with zero, and the $\Gamma$ posterior places $\Gamma = 0$ near the edge of the 90\% credible interval. Unlike for GW200105\_162426, the joint posterior is compatible with $\Gamma = 0$ and $e = 0$ simultaneously at the 90\% level. 
The corresponding Bayes factor (Table~\ref{tab_nsbh}) is $\log_{10}\mathcal{B}^{\Gamma\neq 0}_{\Gamma=0} \approx -0.56$: the eccentric $\Gamma = 0$ hypothesis is still preferred over its $\Gamma\neq 0$ extension.
Compared to Ref.~\cite{Pathak:2026cik}, who included a leading-order eccentricity phase correction within \texttt{IMRPhenomXPHM}, the two-dimensional $(e, \Gamma)$ posterior shows no strong LOSA--eccentricity correlation, likely reflecting the more detailed modeling of eccentricity in \texttt{SEOBNRv6EHM} beyond the leading-order phase correction. On the one-dimensional marginals, Ref.~\cite{Pathak:2026cik} obtains $\Gamma \approx -2.9 \times 10^{-3}\,\mathrm{s}^{-1}$ and $e_0 \approx 0.11$, with the 90\% credible interval on $\Gamma$ excluding zero, whereas our analysis yields a $\Gamma$ posterior peaked closer to zero ($\Gamma \approx -1.04 \times 10^{-3}\,\mathrm{s}^{-1}$) and an eccentricity consistent with zero. Both analyses concur that the data do not provide statistically significant evidence for non-zero LOSA or non-zero eccentricity.

A complementary \texttt{SEOBNRv5PHM} analysis (Fig.~\ref{fig:res_corners_190814}) recovers a $\Gamma$ posterior peaked at $\Gamma \sim -2 \times 10^{-3}\,\mathrm{s}^{-1}$, with $\Gamma = 0$ slightly outside the 90\% credible interval and $\log_{10}\mathcal{B}^{\Gamma\neq 0}_{\Gamma=0} \approx 0.36$---a weak preference for non-zero LOSA, broadly agreeing with the mode-consistent \texttt{IMRPhenomXPHM} analysis of Ref.~\cite{Roy:2026mco}. The Bayes factor between the two $\Gamma\neq 0$ analyses, computed from the nested-sampling evidences, is $\log_{10}\mathcal{B}^{\texttt{v6EHM},\,\Gamma\neq 0}_{\texttt{v5PHM},\,\Gamma\neq 0} \approx -0.28$, mildly favoring the precessing template. Combined with the \texttt{SEOBNRv6EHM} value above, this yields
\begin{equation*}
\log_{10}\mathcal{B}_{\texttt{v6EHM},\,\Gamma = 0}^{\texttt{v5PHM},\,\Gamma\neq 0} \approx - 0.56 + 0.28 \approx - 0.28,
\end{equation*}
indicating that the eccentric non-accelerating hypothesis is mildly preferred over the precessing accelerating alternative once the Occam penalty for the extra $\Gamma$ degree of freedom is taken into account. Within the eccentric template, the corresponding Bayes factor between the $e \neq 0$ and $e = 0$ hypotheses is $\log_{10}\mathcal{B}^{e\neq 0}_{e=0} \approx -1.19$~\cite{Pompili:2026yxq}: neither $\Gamma$ nor $e$ is individually preferred, and the most natural overall interpretation of GW190814 remains a quasi-circular, non-accelerating source.

\renewcommand{\arraystretch}{1.6}
\begin{table}[t]
\centering
\small
\begin{tabular*}{\columnwidth}{@{\extracolsep{\fill}} l c c}
\hline\hline
$\log_{10}\mathcal{B}^{\mathcal{H}_A}_{\mathcal{H}_B}$ & GW200105\_162426 & GW190814 \\
\hline
\texttt{v6EHM}, $\Gamma\neq 0$ \;vs\; \texttt{v6EHM}, $\Gamma=0$ & $-0.60^{+0.02}_{-0.03}$ & $-0.56^{+0.04}_{-0.04}$ \\
\texttt{v5PHM}, $\Gamma\neq 0$ \;vs\; \texttt{v6EHM}, $\Gamma=0$ & $-0.94^{+0.10}_{-0.10}$ & $-0.28^{+0.13}_{-0.13}$ \\
\texttt{v6EHM}, $\Gamma\neq 0$ \;vs\; \texttt{v5PHM}, $\Gamma=0$ & $0.80^{+0.24}_{-0.21}$ & $0.09^{+0.28}_{-0.20}$ \\
\texttt{v5PHM}, $\Gamma\neq 0$ \;vs\; \texttt{v5PHM}, $\Gamma=0$ & $0.45^{+0.21}_{-0.18}$ & $0.36^{+0.25}_{-0.15}$ \\
\hline\hline
\end{tabular*}
\caption{$\log_{10}$ Bayes factors between the $\Gamma \neq 0$ and $\Gamma = 0$ hypotheses for GW200105\_162426 and GW190814, across the \texttt{SEOBNRv6EHM} (\texttt{v6EHM}; eccentric aligned-spin) and \texttt{SEOBNRv5PHM} (\texttt{v5PHM}; precessing quasi-circular) templates. For GW200105\_162426 the eccentric non-accelerating hypothesis is preferred; the positive support for $\Gamma \neq 0$ with \texttt{SEOBNRv5PHM} reflects the eccentricity-induced spurious LOSA bias seen on the EAS injection (Sec.~\ref{sec:val_eas}). For GW190814 no significant LOSA evidence is found under either template; combined with the disfavoring of $e \neq 0$ with \texttt{SEOBNRv6EHM}, the data overall favor a quasi-circular, non-accelerating source.}
\label{tab:bfs}
\end{table}

\subsubsection{Astrophysical implications}

The marginal $\Gamma$ posteriors of Table~\ref{tab_nsbh} can be recast as exclusion regions on the mass $M_\bullet$ and separation $r$ of a putative tertiary perturber via Eq.~(6) of Ref.~\cite{Vijaykumar:2023tjg},
\begin{equation}
\label{eq:perturber_Gamma}
\Gamma \approx 4.65 \times 10^{-12}\,\mathrm{s}^{-1}
\left(\frac{M_\bullet}{10^{10}\,M_\odot}\right)
\left(\frac{r}{1\,\mathrm{pc}}\right)^{-2}
\cos\theta,
\end{equation}
where $\theta$ is the angle between the acceleration vector (which, for a spherically symmetric potential, is aligned with the perturber--binary separation) and the line of sight, so the LOSA is largest when the acceleration points along the line of sight ($\cos\theta = \pm1$, the binary accelerating directly toward or away from us) and vanishes when it is transverse to it ($\cos\theta = 0$). For illustration, the tightest bound in our sample, $|\Gamma| \lesssim 2.4 \times 10^{-4}\,\mathrm{s}^{-1}$ for GW230529\_181500, would exclude a $10^{9}\,M_\odot$ SMBH companion at $r \lesssim 10\,\mathrm{AU}$ or a $10^{6}\,M_\odot$ BH at $r \lesssim 0.3\,\mathrm{AU}$ for a perturber along the line of sight ($\cos\theta = 1$).

For typical perturber configurations of most compact-binary formation channels, the expected $\Gamma$ lies several orders of magnitude below our current sensitivity, so the non-detection of LOSA in our NSBH sample is expected. Predicted $\Gamma$ distributions peak at $10^{-17}$--$10^{-14}\,\mathrm{s}^{-1}$ for binaries in globular clusters~\cite{Tiwari:2023cpa} and at $\sim 10^{-15}$--$10^{-12}\,\mathrm{s}^{-1}$ for binaries orbiting the central SMBH in galactic nuclei~\cite{Inayoshi:2017hgw, Vijaykumar:2023tjg}. An exception could be mergers in AGN disks, a fraction of which form through binary--single encounters or GW captures in which a third body remains bound to the merging binary at small separation; the acceleration from this close third body, rather than from the distant central SMBH, can be large enough that the upper tail of the resulting GW phase-shift distribution approaches current or near-future ground-based sensitivity, with $\Gamma$ reaching $\mathcal{O}(10^{-4})\,\mathrm{s}^{-1}$~\cite{Tagawa:2025tfd, Zwick:2025wkt}, although these predictions remain model-dependent.
Current LOSA bounds should thus be interpreted primarily as a search for outliers---compact-binary environments more extreme than predicted by standard formation models. 

Notably, the constant-acceleration ansatz is accurate for the near-zero $\Gamma$ probed by our null bounds, but might no longer suffice for an actual detection of such an outlier: on a circular outer orbit, a tertiary producing $\Gamma \sim 10^{-4}\,\mathrm{s}^{-1}$ has a period $P = 2\pi (G M_\bullet)^{1/4}(\Gamma c)^{-3/4} \sim 9\,\mathrm{min}$ for a $10\,M_\odot$ perturber, so the long-duration, low-mass signals best suited to LOSA inference, which span $32$--$128\,\mathrm{s}$ in band for the events analyzed here, would cover a non-negligible fraction of the orbit. A confident detection in this regime would therefore require more generic LOSA modeling. Measuring the realistic $\Gamma$ values predicted across formation channels will be possible with future detectors: next-generation ground-based detectors Einstein Telescope and Cosmic Explorer are forecast to reach $\sim 10^{-9}\,\mathrm{s}^{-1}$~\cite{Vijaykumar:2023tjg, Tiwari:2024pvb}, while the long signal durations and low-frequency sensitivity of LISA~\cite{Tamanini:2019usx, Wong:2019hsq, Toubiana:2020drf, Sberna:2022qbn} and of proposed space-based decihertz detectors like DECIGO~\cite{Vijaykumar:2023tjg, Tiwari:2024pvb, Tiwari:2025qqx} reach $\sim 10^{-16}\,\mathrm{s}^{-1}$ for stellar-mass binaries.

\section{Conclusions}
\label{sec:conclusions}

The main result of this work is the first joint inference of LOSA and orbital eccentricity for the NSBH events in the LVK catalog using an eccentric IMR multipolar waveform model, enabled by a time-domain implementation of LOSA corrections in the \texttt{SEOBNR} models.
The implementation is based on a remap of the time-domain strain, and is therefore agnostic to the mode content and morphology of the signal. It does not assume a QCAS orbit to compute the LOSA corrections, does not require a leading-order Taylor expansion in $\Gamma$, and modulates both the phase and the amplitude of the observed strain by default. It applies with no modifications to EAS (\texttt{SEOBNRv6EHM}) and to QCP (\texttt{SEOBNRv5PHM}) models, reduces to existing frequency-domain implementations under the SPA when restricted to the dominant mode of QCAS binaries, and is computationally cheaper for native time-domain waveforms than a mode-by-mode SPA implementation in the frequency domain~\cite{Roy:2026mco}.

We validated the implementation against simulated signals from QCAS, QCP, and EAS binaries (Sec.~\ref{sec:validation}). Recoveries with the same template confirm the consistent recovery of the injected $\Gamma$ in all cases. The complementary recoveries reveal an asymmetric pattern of biases: for the strongly-precessing QCP configuration considered, missing precession in the recovery template does not bias LOSA inference (\texttt{SEOBNRv6EHM} recovers $\Gamma$ unbiased), whereas missing eccentricity can act as a LOSA mimicker (\texttt{SEOBNRv5PHM} on an eccentric EAS injection recovers $\Gamma$ railing at the lower prior boundary $\Gamma = -10^{-2}\,\mathrm{s}^{-1}$ together with a shift in the inferred chirp mass and spins), as also found by Ref.~\cite{Tiwari:2025aec} on BNS injections. This motivates \texttt{SEOBNRv6EHM} as a suitable template for LOSA inference on potentially eccentric sources; using an eccentric template, the $\Gamma$ posterior for an eccentric source is unbiased, but broader relative to the QCP case, reflecting a degeneracy between $\Gamma$ and $e$.

Applied to the five NSBH events in the LVK catalog (Sec.~\ref{sec:res_NSBH}), all five yield $\Gamma$ posteriors within or close to consistency with zero at the 90\% credible level, with Bayes factors $\log_{10}\mathcal{B}^{\Gamma\neq 0}_{\Gamma=0}$ ranging from $-0.56$ for GW190814 to $-2.22$ for GW230529\_181500---in all cases favoring $\Gamma = 0$. To our knowledge, these are the first LOSA constraints for GW200115\_042309, GW230518\_125908, and GW230529\_181500. 
For GW200105\_162426, the joint \texttt{SEOBNRv6EHM} $(\Gamma, e)$ posterior is incompatible with $\Gamma = 0$ and $e = 0$ simultaneously at the 90\% level---a feature also reported by Ref.~\cite{Roy:2026mco}---supporting the eccentricity hints in this event from previous analyses~\cite{Morras:2025xfu, Planas:2025plq, Jan:2025fps, Kacanja:2025kpr, Pompili:2026yxq, Clarke:2026cuw}. A complementary \texttt{SEOBNRv5PHM} analysis recovered spurious support for non-zero $\Gamma$, consistent with the eccentricity-induced bias seen on the EAS injection. Nevertheless, the eccentric non-accelerating hypothesis is preferred both within the eccentric template ($\log_{10}\mathcal{B}^{\Gamma\neq 0}_{\Gamma=0} \approx -0.60$) and against the precessing accelerating alternative ($\log_{10}\mathcal{B}^{\texttt{v5PHM},\,\Gamma\neq 0}_{\texttt{v6EHM},\,\Gamma = 0} \approx -0.94$), and remains the preferred interpretation.
For GW190814, no significant evidence for non-zero LOSA is found under either template; combined with the disfavoring of $e \neq 0$ with \texttt{SEOBNRv6EHM}~\cite{Pompili:2026yxq}, the data overall favor a quasi-circular, non-accelerating source. The four most significant eccentric BBH candidates of Ref.~\cite{Pompili:2026yxq} all yielded uninformative LOSA posteriors, as expected for the higher masses and shorter signal durations of those events.

Several extensions are natural targets for future work. Time-domain models remain computationally expensive for long-duration signals, despite the speed improvements of \texttt{SEOBNRv6EHM} over earlier generations. More systematic injection studies and catalog-scale analyses would benefit from machine-learning-accelerated samplers such as \texttt{DINGO}~\cite{Dax:2021tsq, Dax:2022pxd}, recently extended to support LOSA inference~\cite{RoussopoulosInPrep}.

Because the time-domain remap is agnostic to the underlying dynamics, the implementation transfers without modification to the forthcoming eccentric--precessing model \texttt{SEOBNRv6EPHM}, which will combine eccentricity and spin precession in a single IMR model.
LOSA corrections can also be applied to the accurate BNS waveform model \texttt{SEOBNRv5THM}~\cite{Haberland:2025luz, Haberland:2026xvj}: this would enable unbiased LOSA inference on BNS events, where missing tidal effects can otherwise mimic LOSA~\cite{Tiwari:2025aec}, and where the long signal durations are expected to yield tighter LOSA constraints than for NSBH sources. Beyond accurate vacuum waveforms, LOSA modeling itself will need to be refined for more complex environmental scenarios and higher-SNR signals. For example, if the perturber causing the acceleration is a stellar-mass BH or a massive star close enough to the binary, the constant-acceleration ansatz breaks down; the time-domain formulation could then be extended to higher time-derivatives of the line-of-sight velocity or to a more general time-dependent orbit parametrization~\cite{Sberna:2022qbn, Hendriks:2026kys}. The perturber's Shapiro time delay can be incorporated in the same way, being itself a time-of-flight effect~\cite{Sberna:2022qbn}. All such generalizations can be straightforwardly accommodated within our framework by computing the appropriate $t_{\rm src}(t_{\rm obs})$ for the chosen outer-orbit model, whereas other gravitational effects of the perturber would require a different treatment~\cite{Santos:2025ass}. The same construction can also model the Doppler modulation from the binary's own gravitational recoil, whose center-of-mass velocity builds up through the coalescence; this effect is present for every binary, not only those with a perturber, and is expected to become relevant for LISA and next-generation ground-based detectors~\cite{Gerosa:2016vip, Chamberlain:2018snj}. As future ground-based and space-based detectors approach the sensitivity required for LOSA detections, accurate waveform modeling---both of the vacuum signal and of the environment-induced modulation beyond the constant-acceleration limit---will be essential to translate detections into reliable astrophysical inferences.

While this work was being finalized, an independent, related study appeared~\cite{Roy:2026duh}, which similarly models LOSA as a time-domain Doppler remap and analyzes events across the O1--O4a LVK observing runs.


\section*{Acknowledgments}
We thank Stephen Green, Gonzalo Morr\'as, Alexander Roussopoulos, Laura Sberna, and Aditya Vijaykumar for helpful discussions and feedback, and Laura Sberna in particular for suggesting the time-domain treatment of LOSA effects.
The computational work for this manuscript was carried out on the \texttt{Hypatia} computer cluster at the Max Planck Institute for Gravitational Physics in Potsdam.
L.P. is supported by a UKRI Future Leaders Fellowship (grant number MR/Y018060/1).
A.B and A.G. are supported in part by the European Research Council (ERC) Horizon Synergy Grant “Making Sense of the Unexpected in the Gravitational-Wave Sky” (GWSky-101167314). 
This material is based upon work supported by NSF's LIGO Laboratory which
is a major facility fully funded by the National Science Foundation.
This research has made use of data or software obtained from the Gravitational Wave Open Science
Center (gwosc.org), a service of LIGO Laboratory, the
LIGO Scientific Collaboration, the Virgo Collaboration,
and KAGRA. LIGO Laboratory and Advanced LIGO are
funded by the United States National Science Foundation
(NSF) as well as the Science and Technology Facilities
Council (STFC) of the United Kingdom, the Max-Planck-Society (MPS), and the State of Niedersachsen/Germany
for support of the construction of Advanced LIGO and
construction and operation of the GEO600 detector. Additional support for Advanced LIGO was provided by the
Australian Research Council. Virgo is funded, through
the European Gravitational Observatory (EGO), by the
French Centre National de Recherche Scientifique (CNRS),
the Italian Istituto Nazionale di Fisica Nucleare (INFN)
and the Dutch Nikhef, with contributions by institutions
from Belgium, Germany, Greece, Hungary, Ireland, Japan,
Monaco, Poland, Portugal, Spain. KAGRA is supported
by Ministry of Education, Culture, Sports, Science and
Technology (MEXT), Japan Society for the Promotion
of Science (JSPS) in Japan; National Research Foundation (NRF) and Ministry of Science and ICT (MSIT) in
Korea; Academia Sinica (AS) and National Science and
Technology Council (NSTC) in Taiwan.


\appendix

\bibliographystyle{JHEP}
\bibliography{references}

\end{document}